\documentstyle[twoside,fleqn,espcrc2,epsf]{article}
\renewcommand{\floatsep}{5mm plus 2pt minus 0pt}
\textfloatsep=\floatsep
\intextsep=\floatsep
\newcommand{\pl}{{\rm pl}}
\newcommand{\rt}{{\rm rt}}
\newcommand{\pg}{{\rm pg}}
\title{\bf The scalar glueball from a
 tadpole-improved action}
\author{Colin Morningstar
 and Mike Peardon\address{Department of Physics
 \& Astronomy, University of Edinburgh,  Edinburgh EH9 3JZ, Scotland}}
\begin{document}

\begin{abstract}
The scalar glueball mass and the string tension are computed in
lattice SU(3) gauge theory with the aim of establishing the
effectiveness of the improved action approach in removing
finite-spacing artifacts.
\end{abstract}

\maketitle
\section{INTRODUCTION}
The removal of lattice-spacing artifacts from numerical estimates
of physical quantities is a central problem in lattice field theory.
Taking the continuum limit by decreasing the coarseness of the mesh
and extrapolating is a computationally-expensive task.  The use of
{\em improved actions} promises to be a much more efficient means
of reducing cutoff contamination \cite{alford}.
The purpose of this work is to examine the effectiveness of the
improved action approach in lattice SU(3) gauge theory by
studying the low-lying glueball spectrum.

In this work, we present results for the scalar glueball mass
$m_g$ and the string tension $\sigma$
using the tadpole-improved L\"uscher-Weisz action with lattice
spacings $a\approx 0.25$ and $0.40$ fm.  Comparing to
results from the standard Wilson action, we find a significant
reduction in the finite-spacing errors in the ratio
$m_g/\sqrt{\sigma}$
(setting the scale using the string tension).
The calculation exposes two problems with using the L\"uscher-Weisz
action on a coarse lattice:  first, the coarseness of the lattice in
the temporal direction severely limits the number of statistically-useful
correlator measurements and hampers the demonstration of plateaux
in effective masses; secondly, the unusual properties of the transfer
matrix can reduce the effectiveness of the variational method in
diminishing excited-state contamination.  These properties are
due to the presence in the improved action of terms which couple
fields separated by more than one time slice.

First, the improved action is described in Sec.~\ref{sec:action}.
Simulation details are then presented in Sec.~\ref{sec:simulate},
followed by results and conclusions.

\section{THE IMPROVED GAUGE ACTION}
\label{sec:action}
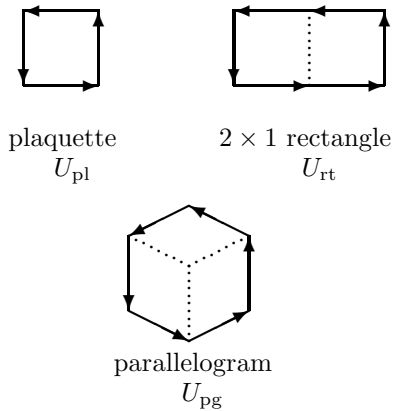
\begin{figure}
\begin{center}
\setlength{\unitlength}{1mm}
\begin{picture}(60,54)
\put(6,44){
   \begin{picture}(10,10)
   \thicklines
   \put(0,0){\vector(1,0){10}}
   \put(10,0){\vector(0,1){10}}
   \put(10,10){\vector(-1,0){10}}
   \put(0,10){\vector(0,-1){10}}
   \put(-2,-8){plaquette}
   \put(4,-12){$U_\pl$}
   \end{picture} }
\put(34,40){
   \begin{picture}(20,18)
   \thicklines
   \put(0,4){\vector(1,0){10}}
   \put(10,4){\vector(1,0){10}}
   \multiput(10,4)(0,1){10}{\circle*{0.3}}
   \put(20,4){\vector(0,1){10}}
   \put(20,14){\vector(-1,0){10}}
   \put(10,14){\vector(-1,0){10}}
   \put(0,14){\vector(0,-1){10}}
   \put(-2,-4){$2 \times 1$ rectangle}
   \put(9,-8){$U_\rt$}
   \end{picture} }
\put(20,10){
   \begin{picture}(16,18)
   \thicklines
   \put(0,4){\vector(2,-1){8}}
   \put(8,0){\vector(2,1){8}}
   \multiput(8,0)(0,1){10}{\circle*{0.3}}
   \multiput(8,10)(1,0.5){9}{\circle*{0.3}}
   \multiput(8,10)(-1,0.5){9}{\circle*{0.3}}
   \put(16,4){\vector(0,1){10}}
   \put(16,14){\vector(-2,1){8}}
   \put(8,18){\vector(-2,-1){8}}
  \put(0,14){\vector(0,-1){10}}
   \thinlines
  \put(-2,-4){parallelogram}
  \put(7,-8){$U_\pg$}
  \end{picture} }

\end{picture}
\end{center}
\vspace{-12mm}
\caption[loops]{
\label{figloops}
Wilson loops in the improved action.}
\end{figure}

The perturbatively-improved QCD gauge action with tadpole improvement
advocated in Ref.~\cite{alford} was used in this study.
The original perturbative improvement was carried out long ago in
Refs.~\cite{LuscherImproved,LuscherOneLoop}; the application
of tadpole improvement \cite{LepMackPert} to this action
is more recent.  The action is given by
\begin{eqnarray}
S[U] &=& \beta_\pl\ \sum_\pl {\textstyle\frac{1}{3}}
          \mbox{Re Tr}\;(1-U_\pl) \nonumber \\
     &+& \beta_\rt\ \sum_\rt {\textstyle\frac{1}{3}}
          \mbox{Re Tr}\;(1-U_\rt) \nonumber \\
     &+& \beta_\pg\ \sum_\pg {\textstyle\frac{1}{3}}
          \mbox{Re Tr}\;(1-U_\pg),
\label{action}
\end{eqnarray}
where $U_\pl$ denotes the usual plaquette, $U_\rt$ indicates
the product of link variables about a planar $2\times 1$ rectangular loop,
and $U_\pg$ is the parallelogram loop.  These terms are depicted in
Fig.~\ref{figloops}.

The coupling $\beta_\pl$ is the only input parameter.  The other
couplings are computed in tadpole-improved perturbation theory
by removing $O(a^2)$ errors in spectral quantities order-by-order
in the QCD coupling.  To one-loop order, these couplings are
given in SU(3) by
\begin{eqnarray}
\beta_\rt &=& -\frac{\beta_\pl}{20 u_0^2}\ (1+0.4805\ \alpha_\Box),\\
\beta_\pg &=& -\frac{\beta_\pl}{u_0^2}\ 0.03325\ \alpha_\Box,
\end{eqnarray}
where the mean-field parameter $u_0$ and the QCD coupling
$\alpha_\Box$ are given in terms of the measured expectation
value of the plaquette as follows:
\begin{eqnarray}
u_0 &=&\langle {\textstyle\frac{1}{3}}{\rm Re Tr}\; U_\pl\rangle^{1/4},\\
\alpha_\Box &=& - 1.30362\ \ln u_0.
\end{eqnarray}
Using identities from Ref.~\cite{LuscherImproved}, the action $S[U]$
was found in Ref.~\cite{alford} to be positive semi-definite for
$\beta_\pl \geq 6.8$.

Since the input couplings depend on the measured quantity $u_0$,
the appropriate value for $u_0$ must be determined by tuning
for each value of $\beta_\pl$.  There are many ways this could be done.
  We chose to fix the $\beta_\pl$ value and vary the input
value of $u_0$ until agreement between the input and measured values
is reached.  An easy and efficient way to do this is to use the
Ferrenberg-Swendsen technique \cite{ferrenberg} to simultaneously
(in a single Markov chain) measure $u_0$ for a range of input values.
These runs are very inexpensive.
A sample of our tuning results follows:
$u_0^4(6.8)=0.467$,  $u_0^4(6.9)=0.480$,  $u_0^4(7.0)=0.494$, and
$u_0^4(7.4)=0.555$.

Due to the presence of terms which span two time slices in the action
$S[U]$ given in Eq.~\ref{action}, the concept of a transfer matrix
is more complicated than for the simple Wilson action.
L\"uscher and Weisz \cite{ImprTransferMatrix} have shown that
a transfer matrix can be defined, but it is in general not
Hermitian; complex eigenvalues can occur, leading to damped
oscillatory behaviour in correlation functions, and negative weight
contributions can appear in the spectral decomposition of two-point
functions.  Nevertheless, the mass spectrum can still be obtained
from the exponential decay of correlators for large
temporal separations.  Care must be exercised, however, in applying
variational methods to extract masses from correlators at short
time intervals:  although invalid, strictly speaking, the variational
method may still be practical whenever the coherence length is
large enough such that the contributions from the
problematic high-energy states are negligible.

\section{SIMULATION DETAILS}
\label{sec:simulate}

A set of $18$ scalar glueball creation operators $\{O_i(t)\}$
were constructed from combinations of Wilson loops on time-slice $t$
with appropriate quantum numbers.
Due to the coarseness of our lattice, fuzzing was not performed.
We then obtained Monte Carlo estimates of the following matrix
of correlators:
\begin{equation}
C_{ij}(\tau) = \sum_t \left\{\langle O_i (t) \; O_j (t\!+\!\tau) \rangle
 -\langle O_i\rangle \; \langle O_j \rangle\right\}.
\end{equation}
The simulations were run mainly on DEC-$\alpha$ and
Hewlett-Packard workstations; 3000 CPU hours of Cray T3D time
were also used.  The T3D was employed essentially as a
task farm consisting of 64 isolated workstations.
The gauge fields were updated using both Cabbibo-Marinari [CM]
pseudoheatbath and overrelaxation [OR] techniques.  Updates
with the improved action were found to be five times more costly
than for the Wilson action.  One CM and one OR sweep were performed
between correlator measurements, and the results were combined
into bins of 1000 to eliminate autocorrelations.  Signals were
enhanced by exploiting Euclidean lattice symmetries.
Results were obtained using $\beta_\pl=6.8$ on a $6^4$ lattice
(334 bins) and $\beta_\pl=7.4$ on an $8^4$ lattice (174 bins)
with periodic boundary conditions.  One simulation using the
Wilson action with $\beta_W=5.5$ was performed
on an $8^4$ lattice (104 bins).

\begin{figure}[t]
\begin{center}
\leavevmode
\setlength\epsfxsize{75mm}
\epsfbox{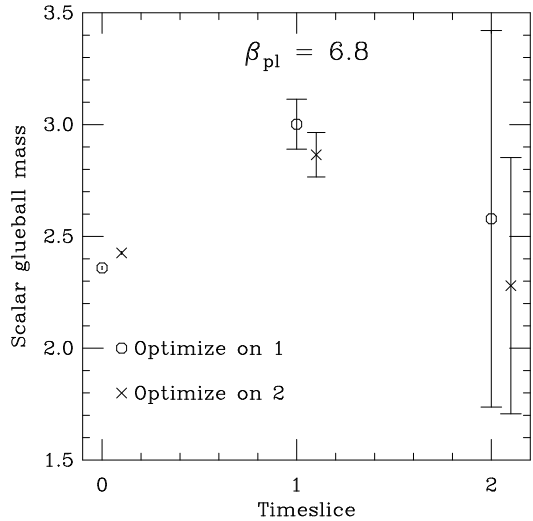}
\vspace{-10mm}
\caption{Effective mass for $\beta_\pl = 6.8$
\label{fig:b68EffMass}}
\end{center}
\end{figure}

To determine $m_g$,
the variational method \cite{BergBilloire} was first used to select
optimal combinations of glueball operators $O_{\rm opt}
= \sum_j c_j O_j$; this involved choosing the $c_j$'s such that
$C_{\rm opt}(t_{\rm opt})/C_{\rm opt}(0)$ was a maximum,
 where $C_{\rm opt}$
is the correlator associated with $O_{\rm opt}$.
The optimizations were performed using
$t_{\rm opt}=1,2$.  The effective masses associated with these
optimal correlators were then examined.
The coarseness of the lattice severely limited
the number of time separations for which signals could be obtained;
for this reason, a sophisticated correlated analysis was not
possible.  Effective masses for glueballs in other channels were
also studied, but no acceptable signals were obtained.

In order to set the scale, we also measured the string tension $\sigma$.
This was done by extracting the static quark potential $V({\bf r})$
for various ${\bf r}$ from ratios of appropriate Wilson loops.
The results were then fit to the form
\begin{equation}
V(r) = \sigma r - \frac{\pi}{12 r} + b,
\end{equation}
in order to obtain $a^2 \sigma$.  Only on-axis ${\bf r}$
vectors were used.

\begin{figure}[t]
\begin{center}
\leavevmode
\setlength\epsfxsize{75mm}
\epsfbox{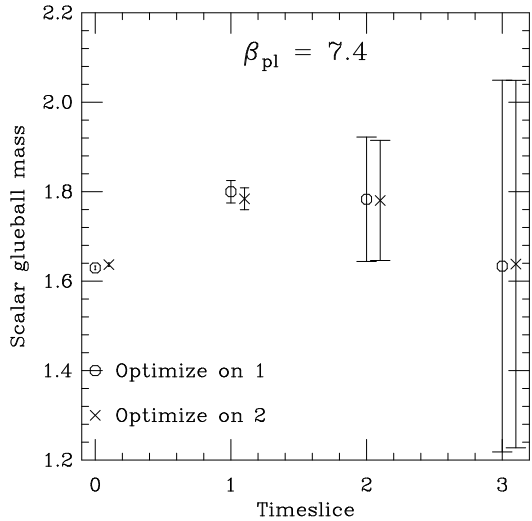}
\vspace{-10mm}
\caption{Effective mass for $\beta_\pl = 7.4$
\label{fig:b74EffMass}}
\end{center}
\end{figure}

\begin{figure}[t]
\begin{center}
\setlength{\unitlength}{1mm}
\begin{picture}(75,75)
\put(0,0){\epsfxsize=75mm \epsfbox{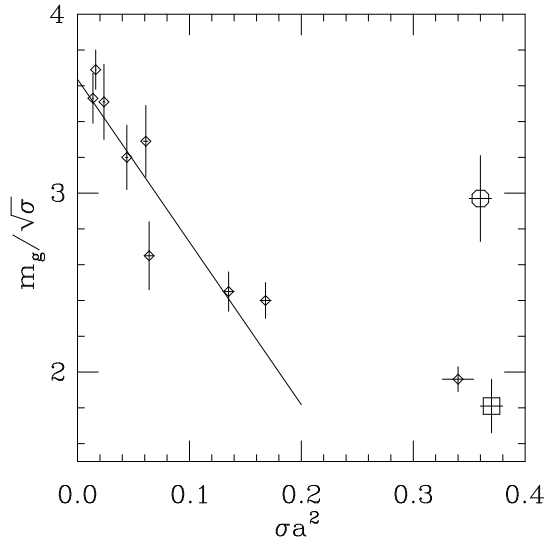}}
\end{picture}
\vspace{-10mm}
\caption[fig5]{Glueball mass against string tension:
 Wilson and improved actions.
The circle is our $\beta_{\rm pl}=7.4$ result; the square indicates
our $\beta_W=5.5$ Wilson action result.  The diamonds are Wilson action
results from Refs.~\cite{Michael,UKQCDGlueballs,Forcrand,Weingarten}.
The straight line is a fit from Ref.~\cite{UKQCDGlueballs} to the
Wilson data.
\label{fig:ScalarGlueball}}
\end{center}
\end{figure}

\section{RESULTS}
\label{sec:results}

Our results for the on-axis string tension are $a^2\sigma(6.8)=0.91(2)$
and $a^2\sigma(7.4)=0.36(1)$.
Systematic uncertainties associated with the form of $V(r)$ used
to extract $\sigma$ dominate over statistical uncertainties;
the errors quoted are estimates of the systematic effects.
Taking $\sqrt{\sigma}=440$ MeV, the lattice spacing can be set,
yielding $a(6.8)=0.427(5)$ fm and $a(7.4)=0.269(4)$ fm.

The effective glueball mass plots are shown in Figs.~\ref{fig:b68EffMass}
and \ref{fig:b74EffMass}.  Circles (crosses) indicate results extracted
from the correlator of the glueball operator obtained by optimizing on
time-slice 1 (2).  Note that the effective masses do not decrease
monotonically; this behaviour is a result of the presence of the
two-time-step interaction in the improved action.  Evidence of a plateau
is clearly visible in the $\beta_\pl=7.4$ plot; large errors shed some
doubt on the existence of a plateau in the $\beta_\pl=6.8$ plot.  It is
interesting to note that the plateau used in Ref.~\cite{Weingarten}
to obtain $m_g$ at $\beta_W=6.4$
lies in the range $0.2\!-\!0.6$ fm,
whereas $t=1\!-\!3$ corresponds to $0.3\!-\!0.8$ fm for our $\beta_\pl=7.4$
results, and $t=1\!-\!2$ corresponds to $0.4\!-\!0.9$ fm for the
$\beta_\pl=6.8$ data.

Finite-volume effects on our glueball masses are expected to be
small.  Using L\"uscher's formula \cite{infvol}, we infer that
our mass estimates lie within $\frac{1}{2}\%$ of their infinite volume
limits.  Note that our volumes $V(6.8)\approx(2.6\ {\rm fm})^4$ and
$V(7.4)\approx(2.2\ {\rm fm})^4$ are of similar size to those used in
Ref.~\cite{Weingarten}.

We are confident that the $t=2$ effective mass value shown
in Fig.~\ref{fig:b74EffMass} gives a
reliable determination of the glueball mass at $\beta_\pl=7.4$.
This value is shown in Fig.~\ref{fig:ScalarGlueball} where it
is compared with glueball determinations from various earlier
works using the Wilson action.
  Our glueball mass using the Wilson action
with $\beta_W=5.5$ is also shown in this plot and
agrees with a previous determination \cite{Forcrand}.  Comparing
the Wilson $\beta_W=5.5$ result with the $\beta_\pl=7.4$ ratio
from the improved action, one observes a remarkable
three-fold reduction in the finite-$a$ errors.

We are less confident that the $t=1$ effective mass shown in
Fig.~\ref{fig:b68EffMass} reliably represents the glueball mass
at $\beta_\pl=6.8$.  However, if we use this value and convert
to physical units using $\sqrt{\sigma}=440$ MeV, we obtain
$m_g(6.8)=1.38(7)$ GeV, compared with $m_g(7.4)=1.31(11)$ GeV.
Results at more $\beta_\pl$ values are needed before one can
safely comment about scaling properties.  Note that the above values
are low in comparison to the $a\rightarrow 0$ Wilson results.

\section{CONCLUSION}
We computed the scalar glueball mass and the string tension
using the tadpole-improved L\"uscher-Weisz SU(3) action on two
coarse lattices and found
a significant reduction in the finite lattice spacing errors in
the ratio $m_g/\sqrt{\sigma}$.  The large lattice spacings in the
temporal direction limited the number of statistically-useful
effective mass measurements.  In the future, we intend to
circumvent this problem by working on asymmetric coarse lattices
in which the spacing in the time direction is reduced.  We expect
that the powerlaw increase in simulation cost will be more
than offset by the exponential increase in signal-to-noise.

We acknowledge discussions with R.~Edwards, G.P.~Lepage,
J.~Mandula, C.~Michael, H.~Trottier, and P.~van Baal, and the financial
support of PPARC, especially through grant GR/J 21347.
We are also grateful to the University of Edinburgh for
generously providing access to the T3D.


\begin{thebibliography}{99}

\bibitem{alford}
M.~Alford, {\it et al.},
Nucl.\ Phys.\ {\bf B} (Proc.\ Suppl.) {\bf 42}, 787 (1995);
FERMILAB-Pub 95/199-T, hep-lat/9507010.

\bibitem{LuscherImproved}
M.~L\"uscher and P.~Weisz,
Comm.\ Math.\ Phys {\bf 97} 59 (1985).

\bibitem{LuscherOneLoop}
M.~L\"uscher and P.~Weisz,
Phys.\ Lett.\ B158, 250 (1985).

\bibitem{LepMackPert}
  G.P.~Lepage and P.~Mackenzie, Phys.\ Rev.~{\bf D48}, 2250 (1993).

\bibitem{ferrenberg}
A.~Ferrenberg and R.~Swendsen, Phys.\ Rev.\ Lett.~{\bf 61}, 2635 (1988).

\bibitem{ImprTransferMatrix}
M.~L\"uscher and P.~Weisz,
Nucl.\ Phys.\ B {\bf 240}, 349 (1984).

\bibitem{BergBilloire}
  B.~Berg and A.~Billoire, Nucl.\ Phys.~{\bf B221}, 109 (1983).

\bibitem{Weingarten}
H.~Chen, {\it et al.},
Nucl.\ Phys.\ B (Proc.\ Suppl.) {\bf 34}, 357 (1994).

\bibitem{infvol}
M.~L\"uscher, Comm.\ Math.\ Phys.~{\bf 104}, 177 (1986).

\bibitem{Michael}
C.~Michael and M.~Teper, Phys.\ Lett.~{\bf 206B}, 299 (1988);
 Nucl.\ Phys.~{\bf B314}, 347 (1989).

\bibitem{UKQCDGlueballs}
G. Bali {\it et al.}, Phys.\ Lett.~{B309}, 378 (1993).

\bibitem{Forcrand}
P.~de~Forcrand, {\it et al.}, Phys.\ Lett.~{\bf 152B}, 107 (1985);
Phys.\ Lett.~{\bf 160B}, 137 (1985).

\end{thebibliography}
\end{document}